%% file: master.tex
\documentclass[10pt,journal,compsoc]{IEEEtran}
\ifCLASSOPTIONcompsoc
  \usepackage[nocompress]{cite}
\else
  \usepackage{cite}
\fi
\ifCLASSINFOpdf
\else
\fi
\usepackage{amsmath}
\usepackage{graphicx}
\usepackage{amssymb,amsmath}
\usepackage{url}

\begin{document}

%\title{Broker Distribution for the IoT and Supply Chain Monitoring using the Blockchain Technology}
\title{Trinity: A Distributed Publish/Subscribe Broker with Blockchain-based Immutability}


\author{Gowri Sankar Ramachandran,
        Kwame-Lante Wright,
        and~Bhaskar Krishnamachari,~\IEEEmembership{Senior Member,~IEEE}%  <-this % stops a space
\IEEEcompsocitemizethanks{\IEEEcompsocthanksitem G.S. Ramachandran, K.L. Wright and B. Krishnamachari are with the Department of Electrical Engineering, University of Southern California, Los Angeles,
CA, 90089.\protect\\
E-mail: gsramach@usc.educ}
\thanks{Manuscript submitted June 3, 2018.}}

\IEEEtitleabstractindextext{%
\begin{abstract}
Internet-of-Things (IoT) and Supply Chain monitoring applications rely on messaging protocols for exchanging data. Contemporary IoT deployments widely use the publish-subscribe messaging model because of its resource-efficiency. However, the systems with publish-subscribe messaging model employ a centralized architecture, wherein the data from all the devices in the application network flows via a central broker to the subscribers. Such a centralized architecture make publish-subscribe messaging model susceptible to a central point of failure. Besides, it provides an opportunity for the organization that owns the broker to tamper with the data. In this work, we contribute Trinity, a novel distributed publish-subscribe broker with blockchain-based immutability. Trinity distributes the data published to one of the brokers in the network to all the brokers in the network. The distributed data is stored in an immutable ledger through the use of the blockchain technology. Furthermore, Trinity executes smart contracts to validate the data before saving the data on the blockchain. Through the use of a blockchain network, Trinity can guarantee persistence, ordering, and immutability across trust boundaries. Our evaluation results show that Trinity consumes minimal resources, and the use of smart contracts enable the stakeholders to automate the data management processes. To the best of our knowledge, Trinity is the first framework that combines the components of the blockchain technology with the publish-subscribe messaging model. 
\end{abstract}

}

\maketitle

% is not selected <OR> if conference mode is selected - because compsoc
% conference papers position the abstract like regular (non-compsoc)
\IEEEdisplaynontitleabstractindextext

% \fi
\IEEEpeerreviewmaketitle

%\ifCLASSOPTIONcompsoc
%\IEEEraisesectionheading{\section{Introduction}\label{sec:introduction}}
%\else
%\section{Introduction}
%\label{sec:introduction}
%\fi
\input{Introduction}
\input{Blockchain_Architecture}

\input{motivation}
\input{system}

\input{results}

\section{Related Works}
\input{related_works}

\input{conclusion}

% or


% \appendices

% % you can choose not to have a title for an appendix
% % if you want by leaving the argument blank
% \section{}

\ifCLASSOPTIONcompsoc
  \section*{Acknowledgments}
\else
  \section*{Acknowledgment}
\fi
This work is supported by the USC Viterbi Center for Cyber-Physical Systems and the Internet of Things (CCI).

\ifCLASSOPTIONcaptionsoff
  \newpage
\fi



%\bibliographystyle{IEEEtran}
%\clearpage
\bibliographystyle{IEEEtran}
\bibliography{references}


% H.~Kopka and P.~W. Daly, \emph{A Guide to {\LaTeX}}, 3rd~ed.\hskip 1em plus

% \end{thebibliography}


% \end{IEEEbiography}


%\newpage



%\vfill


\end{document}

%% file: Introduction.tex
\section{Introduction}
IoT applications use typically publish-subscribe\cite{mqtt} and request-reply\cite{shelby2014constrained} messaging models for exchanging data between end-devices, edge devices, and servers. The request-reply messaging model is widely used on the Internet by HTTP protocol. The request-reply messaging model is well tested and standardized in the Internet context, but it is not ideally suitable for resource-constrained IoT systems. CoAP~\cite{shelby2014constrained} is a lightweight request-reply protocol targeted for resource-constrained IoT systems, which consumes limited resources, but it still lacks support for scalability and portability across a wide range of platforms. Alternatively, the publish-subscribe messaging model is widely used because of its low communication overhead and resource efficiency. A number of publish-subscribe systems are proposed in the literature for the IoT such as MQTT~\cite{banks2014mqtt}, LooCI~\cite{Hughes:2009:LLC:1821748.1821787}, NesC~\cite{Levis:2004:ENA:1251175.1251176}, and RUNES~\cite{1651554}.

\begin{figure}[b]
\begin{center}
    \centerline{\includegraphics[height=3in,width=3in, keepaspectratio]{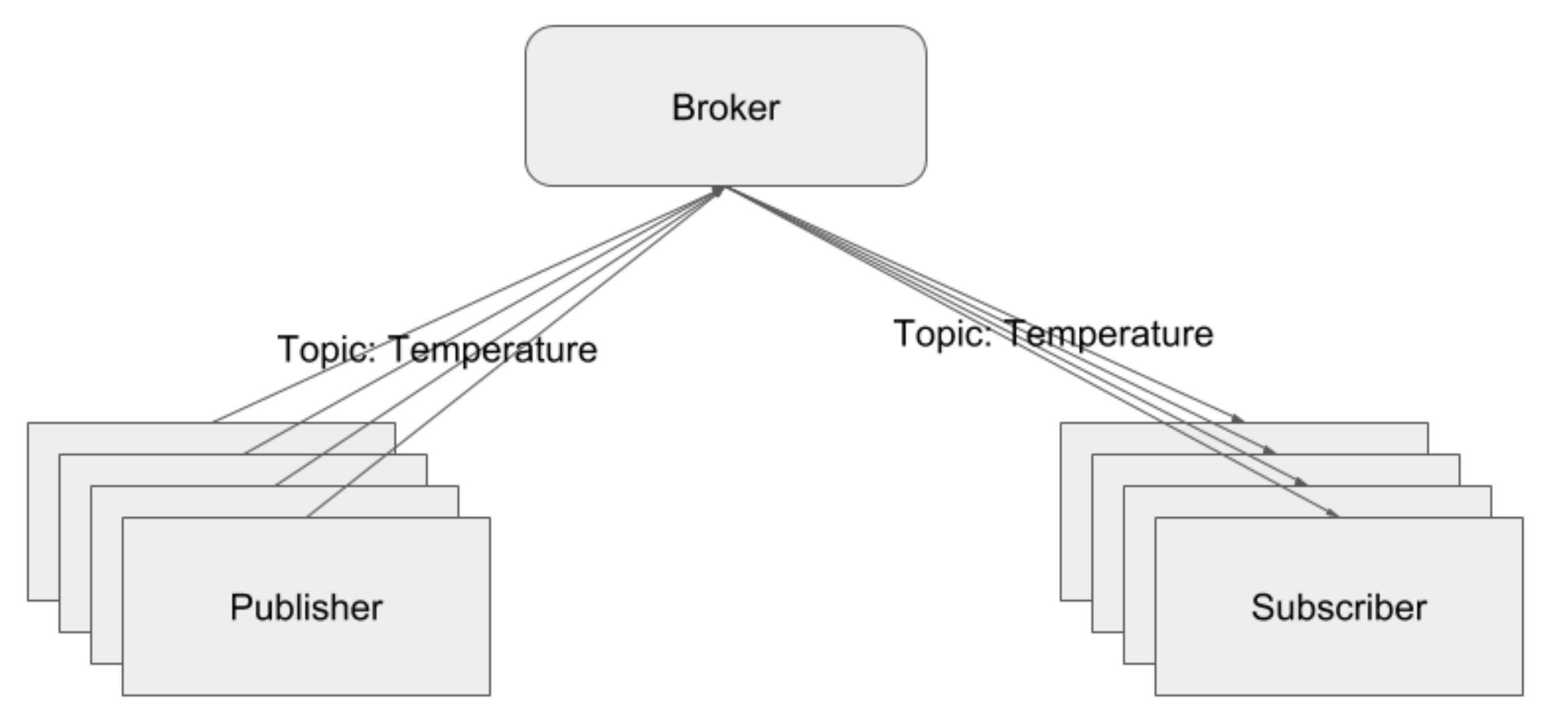} }
    \caption{Publish-Subscribe Communication Model.}
    \label{fig:broker}
\end{center}
\end{figure}

Figure~\ref{fig:broker} shows the interaction model of the publish-subscribe messaging model.
In the publish-subscribe system, publishers and subscribers interact via a broker. A broker is a centralized software that orchestrates the communication between the publishers and subscribers. This messaging model is widely employed in IoT systems in the context of smart cities, industrial IoT, and supply chain monitoring applications because of its resource-efficiency and scalability.

Although the publish-subscribe messaging model is lightweight, scalable, and resource-efficient, it relies on a central broker for data communication between publishers and subscribers. Such a centralized architecture makes publish-subscribe messaging model vulnerable to central points of failure. Besides, it enables a publisher and a subscriber to interact through a central server owned by a single organization. In supply chain monitoring applications, products transit through infrastructure owned and managed by various organizations, and at each organization's warehouse, the status of the assets are monitored, and the state of the product is stored in a central system owned by an individual entity. Any mishandling of products or a lack of maintenance makes products unusable, which might result in a loss for one or more organizations. In such circumstances, it is hard to identify which organization mishandled the asset and approaches based on a single central server would enable an organization to manipulate or erase the data from the central broker or the server.

In this work, we contribute Trinity, a novel distributed publish-subscribe broker with blockchain-based immutability by integrating the broker system with a blockchain framework. The blockchain allows the broker data to be replicated across multiple brokers using a consensus algorithm. Besides, Trinity records all the data in an immutable ledger. Furthermore, our framework provides support for smart contracts, which is used for validating the published data against the agreement made by all the parties involved in the transaction. Whenever a publisher violates the agreement, it is not only registered in the blockchain ledger but also notified to all the stakeholders via the brokers. The Trinity framework is implemented using MQTT broker~\cite{mqtt} and Tendermint~\cite{buchman2016tendermint} blockchain framework. The evaluation results show that the Trinity framework introduces timing and performance overhead to provide assurance when transacting in a multi-stakeholder environment. 

Section 2 discusses the components of the blockchain and the broker in detail. Section 3 introduces the architectural elements of our broker. Section 4 presents the evaluation of our broker framework. Related work is presented in Section 5. Section 6 concludes the paper with the pointers for the future work.

%% file: Blockchain_Architecture.tex
\section{Overview of the blockchain and publish-subscribe messaging model}
The Trinity framework combines the blockchain technology with the publish-subscribe messaging model, which is one of the widely used messaging models in IoT deployments. Section~\ref{sec:overview_blockchain} provides the overview of the blockchain technology, while the publish-subscribe messaging model is discussed in Section~\ref{sec:pubsub}.

\subsection{Blockchain Technology}
\label{sec:overview_blockchain}
The key components of the blockchain technology include consensus algorithm, distributed ledger, and public-key cryptography. The above components communicate and coordinate over a distributed network of devices owned and maintained by multiple entities. If the members involved in the blockchain network are already known to the network, then the blockchain is called the \emph{permissioned} blockchain. When the system is open to the general public, any individual or an organization can be a member of the blockchain network, and such a blockchain is referred to as \emph{public} blockchain. Bitcoin and Ethereum are the examples of a public blockchain, while HyperLedger Fabric and Ripple are examples of permissioned blockchains. The building blocks of blockchain are discussed below:

\subsubsection{Consensus Protocol} 
Consensus protocol is the key component of the blockchain, as it allows a collection of entities to agree on the state of the system. In the context of cryptocurrencies, the term \emph{state} refers to the account balance, and whenever a user initiates a transaction, the members in the network verifies the account to make sure that the user is whom he/she claims to be and has sufficient balance in the account before performing the transaction. All the members of the network maintain the copy of a state at all times. Some of the well-known consensus algorithms in the literature include Proof-of-Work, Proof-of-Stake, BFT, and Raft.

%For our work, a consensus algorithm is necessary to make sure that the members of the network validate the data received at a particular broker before the subscribers receive the data. In broker-driven applications, the consensus layer adds additional trust and the publishers are obliged to act in a benevolent manner. Besides, the subscribers can confidently consume the data as multiple entities approve it. To validate and approve the data in a distributed setting, a consensus algorithm is necessary.  

\subsubsection{Distributed Storage or Ledger}
The blockchain systems follow a decentralized architecture, wherein all the members of the network collectively accomplish the desired application goal. A system state perceived in one device is replicated to all the other devices in the network through the execution of consensus logic and peer-to-peer networking protocol. The replicated state information is stored in a blockchain, which is jointly managed by the members of the network. All the members of the blockchain network uniformly run the blockchain application following the predefined consensus protocol and the block creation protocols. Block in the context of blockchain refers to a distributed storage schema or a ledger. Each block in the blockchain consists of one or more transactions, signature of the block validators, reference to the previous block along with block headers.  

\subsubsection{Public Key Cryptography}
The identity of the users is protected through public-key cryptography in blockchain networks. This cryptography is a type of asymmetric cryptography, wherein a transaction between a sender and a receiver is digitally secured through a pair of keys; private and public keys. The sender signs the transaction with a secret private key, which can only be decrypted by a receiver with a valid public key.

\subsubsection{Smart Contracts}
The smart contract allows the members of the network to validate the state information. Smart contracts consist of a software program that gets triggered by the occurrence of transactions. Whenever a transaction triggers a smart contract, the software logic associated with the contract is executed to carry out certain actions such as checking whether the state of the system is valid or in the case of cryptocurrencies, it might be a payment of cryptocurrency to a member of the network. Smart contracts help the application developers to automate the application processes.

%In the case of supply chain monitoring application as the consumers/subscribers may only keep a copy locally on their broker without actually checking the state of the good. To reduce the maintenance and tracking effort, the product manufacturers agree on a state of the good throughout the supply chain process, and at all transit points throughout the supply chain, the state of the product is compared against the state prescribed in the smart contract. When the product starts deteriorating or if one of the transit points fail to enter the state of the product, all the stakeholders in the supply chain process are notified, and only the transit points and stakeholders involved in the process until the deterioration point is held accountable for the loss. This allows the supply chain companies to track products reliably and efficiently. 

\subsection{Publish-Subscribe Messaging Model}
\label{sec:pubsub}
%Distributed applications rely on a messaging model for coordinating and collaborating with other systems in the network. A wide-array of messaging models were proposed for distributed interactions including RPC, CoAP, and MQTT. Remote Procedure Calls (RPC) allows a program or a function in one machine to interact with a program or a function in other machines in the network without knowing the details of the network protocols. Although RPC offered a powerful approach for distributed communication, it is not suitable for IoT applications because of its high resource and communication overhead. Protocols such as CoAP and MQTT were created for resource-constrained IoT devices. On the one hand, CoAP follows the request-reply communication model, wherein the interested party has to request the data source for the new data periodically. MQTT, on the other hand, follows a publish-subscribe messaging model, in which a broker is used to orchestrate the communication between the data producers and consumers. Even though both MQTT and CoAP are suitable for IoT applications, CoAP is not widely used because of the bidirectional communication requirement for the request and reply, whereas MQTT requires one request to be submitted to a broker, after which the data from the producers are forwarded to the subscribers by the broker. We consider MQTT and publish-subscribe messaging model due to its resource-efficiency and scalability.

As shown in Figure~\ref{fig:broker}, the publish-subscribe communication model typically consists of three components; broker, publisher, and subscriber. Publishers in the system send data to a broker following the concept of the topic. Topic typically refers to the metadata, which describes information about the data in a string format. A topic can have multiple levels. For example, the data generated by a temperature sensor deployed at room 123 of building A can have its topic defined as \emph{\textbackslash  buildingA \textbackslash room123 \textbackslash Temperature}. Consumers of the data can receive data from the temperature sensor by subscribing to the \emph{\textbackslash  buildingA \textbackslash room123 \textbackslash Temperature} topic. In the next section, we discuss the need for a distributed publish-subscribe broker and the blockchain.

%% file: motivation.tex
\section{Why Distributed Broker?}
We will motivate the need for a distributed broker with blockchain-based immutability in this section using a supply-chain monitoring use case. 

%https://www.sciencedirect.com/science/article/pii/S1474034611000553 (supply chain with a single central database)

%actors in food supply chain (https://www.koganpage.com/media/project_kp/document/food-supply-chain-management-and-logistics-sample-chapter.pdf)

\begin{figure}[b]
\begin{center}
    \centerline{\includegraphics[height=3in,width=3in, keepaspectratio]{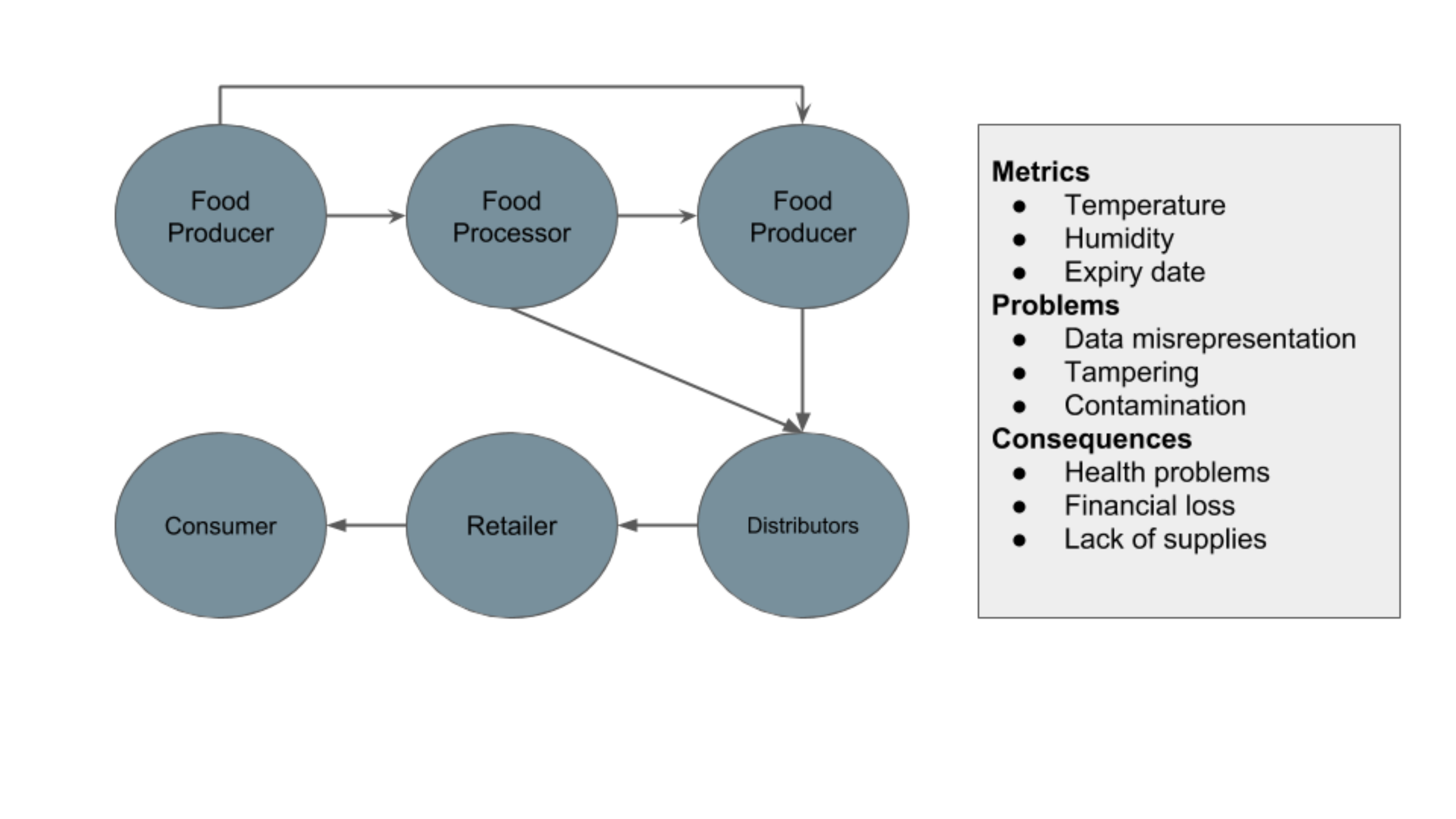} }
    \caption{Food Supply Chain Use Case.}
    \label{fig:usecase}
\end{center}
\end{figure}

\subsection{Food Supply Chain Use Case}
\label{sec:usecase}
The food products that we consume day-to-day passes through multiple storage houses, processing units, and distributors. Figure~\ref{fig:usecase} illustrate the typical supply chain process for food products~\cite{dani2015food}. The food product has to be preserved at the recommended climatic conditions throughout the supply chain to reduce the risk of contaminations. The metrics such as temperature, humidity, and the expiry dates are to be measured to ensure that the product is in a edible state. By gathering and storing this information in a ledger, each organization can ensure that the non-edible food products do not reach both the consumers and the other organizations in the supply chain process. Besides, the organizations that mishandled the food product must be identified to impose a penalty. In all cases, it is important to monitor the status of the product throughout the supply chain to ensure safety as the contaminated food product may cause health issues to a customer, as noted in Figure~\ref{fig:usecase}. The state-of-the-art approaches typically register the status of the food products on the infrastructure owned by each organization in the supply chain~\cite{TRIENEKENS201255}. Even the centralized solutions stores the supply chain information in an infrastructure owned by a single organization. Although such approaches enable the stakeholders in the supply chain to ensure food safety, it does not provide the complete trust since the organizations can tamper with the data to hide their faults. We believe a distributed broker with blockchain-based immutability allows all the organizations in the supply chain process to transact in a trustworthy manner.

\subsection{Single Broker vs. Multiple Brokers}
When a single broker (see Figure~\ref{fig:broker}) is used for the supply chain use case presented in Figure~\ref{fig:usecase}, all the organizations in the supply chain network has to publish the state of the food products to the centralized broker. The owner of the broker can manipulate the data to make the food product appear in a edible state, when it is contaminated or be starting to be spoiled. An alternative solution would be to publish the data to brokers owned by all the organization in the supply chain. Since all the brokers in the network have a copy of the data, it is almost impossible to manipulate the data after it is published. Note that the labors handling the data collection process may enter incorrect data, but we assume that the data collected at the checkpoints are accurate, for this work.

% \begin{figure*}%
%     \centering
%     \subfloat[Multiple Devices Connecting to a Single Broker]{{\includegraphics[width=8cm]{Single_Broker.png} }}%
%     \qquad
%     \subfloat[Multiple Devices Connecting to Multiple Broker]{{\includegraphics[width=8cm]{Multiple_Broker.png} }}%
%     \caption{Single Broker vs Multiple Brokers}%
%     \label{fig:broker_architecture}%
% \end{figure*}

%% file: system.tex
\section{Architecture of Trinity}

\begin{figure}[t]
\begin{center}
    \centerline{\includegraphics[height=3in,width=3in, keepaspectratio]{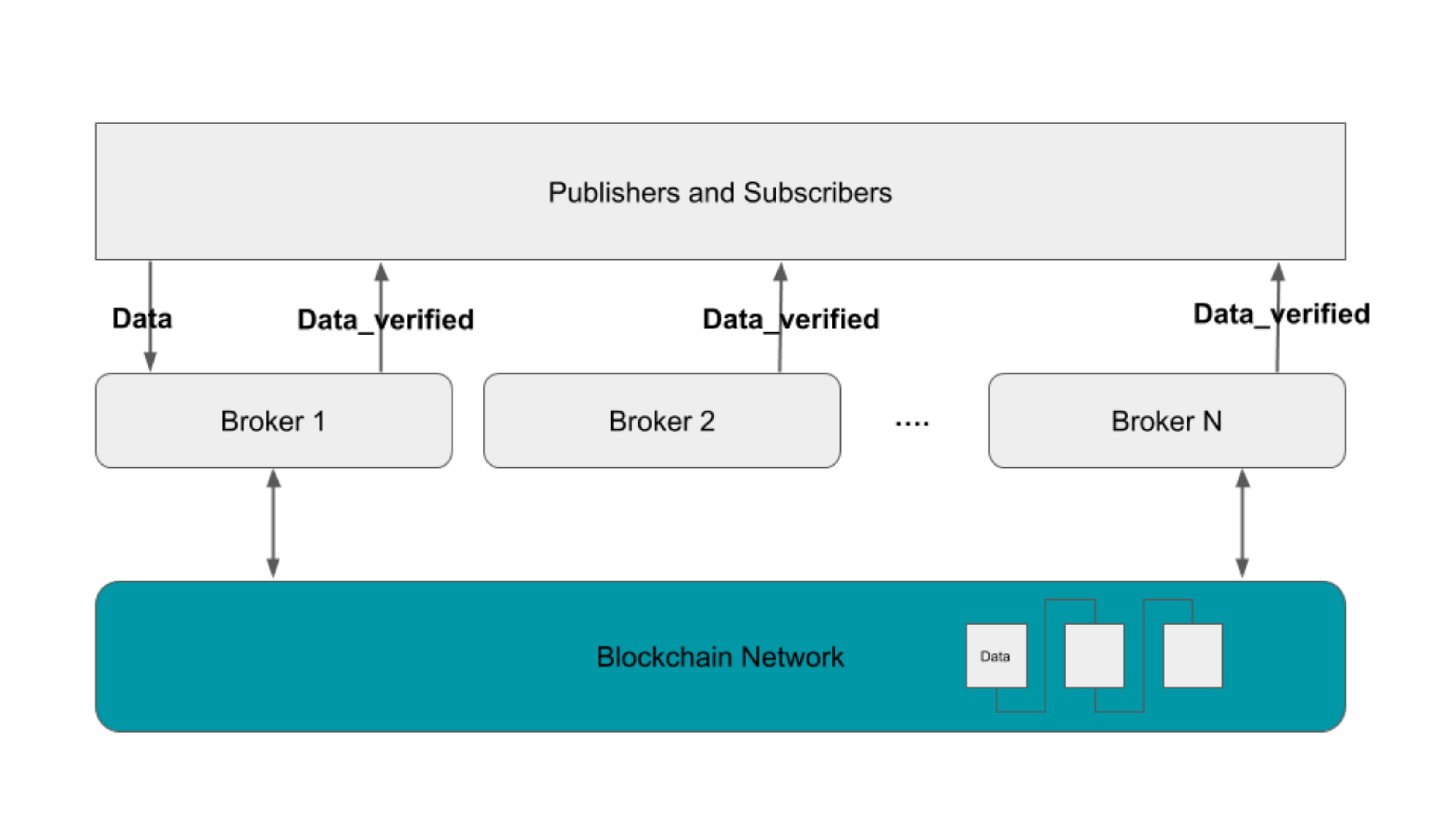} }
    \caption{Overview of the Trinity framework.}
    \label{fig:architecture}
\end{center}
\end{figure}

\begin{figure}[b]
\begin{center}
    \centerline{\includegraphics[height=3in,width=3in, keepaspectratio]{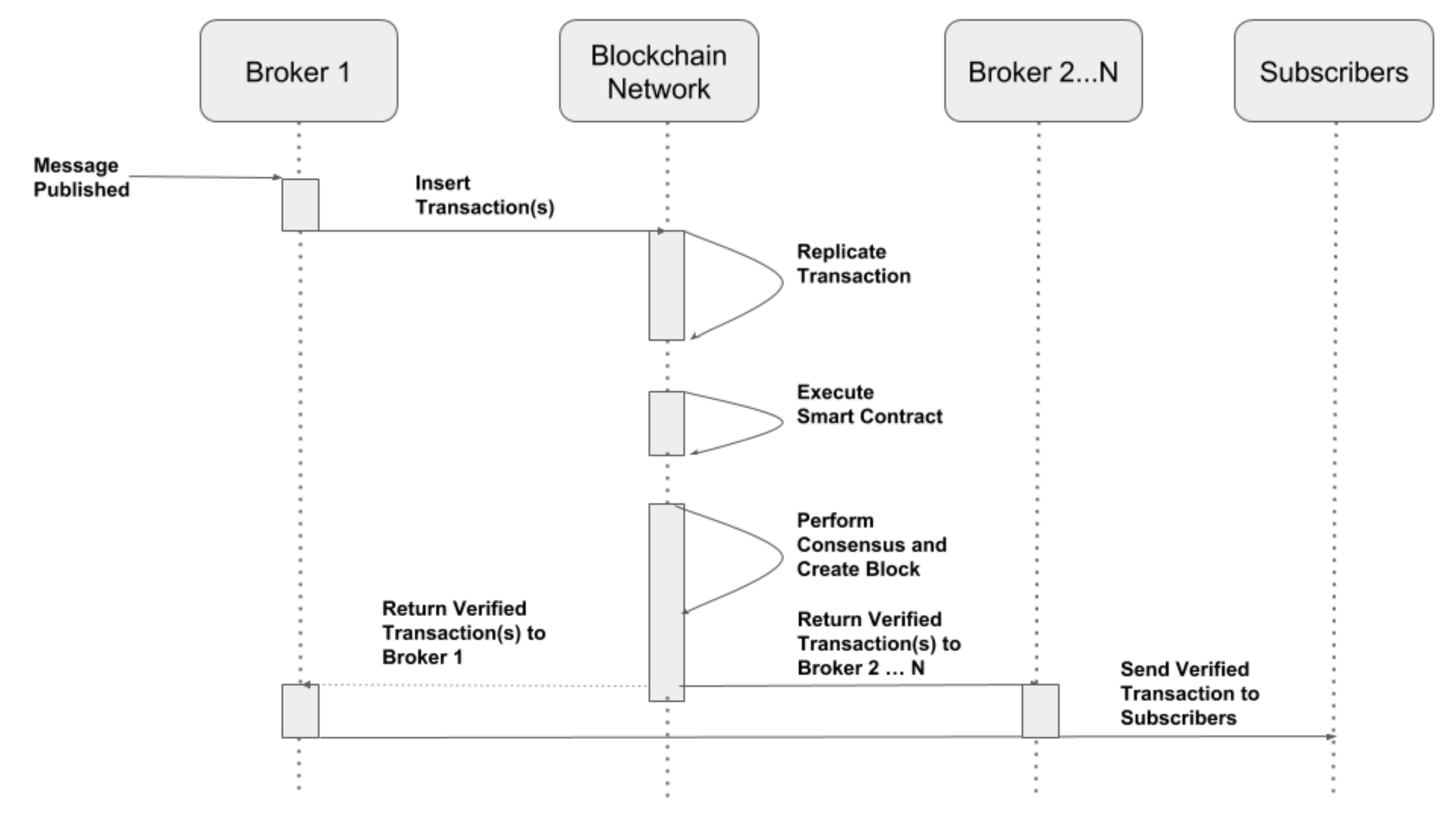} }
    \caption{Sequence diagram showing how the published data is verified in Trinity network.}
    \label{fig:sequence}
\end{center}
\end{figure}

The Trinity framework combines the publish-subscribe broker with the components of the blockchain technology such as distributed ledger and consensus algorithm. Figure~\ref{fig:architecture} shows the main components of Trinity. The three main components of our system include the blockchain network, broker, and publishers and subscribers. All the components are distributed across the network. For the use case presented in Figure~\ref{fig:usecase}, each stakeholder in the supply chain deploy a Trinity instance in their warehouse. The Trinity deployment enables each stakeholder to publish information not only to their brokers, but also to other brokers via the blockchain framework. Figure~\ref{fig:sequence} shows the execution sequence of Trinity, it is explained below for the use case presented in Figure~\ref{fig:usecase}:

\begin{itemize}
\item{All the stakeholders in the supply chain process jointly create a smart contract by including information about the recommended conditions and the topics associated with the contract.}
\item {When a product leaves the warehouse of a distributor, a warehouse manager collects information such as temperature, humidity, and expiry date and publishes the information to his local Trinity broker instance under a topic defined in the contract.}
\item{The Trinity instance receives the published message and check against the list of registered topics to check whether the topic has an associated smart contract in the Trinity network. If the topic has an associated contract, then all the Trinity validators (discussed in Section~\ref{sec:blockchain_net}) must execute the smart contract and come to a consensus on the state of the good. In all other cases, the \emph{unverified} transaction is sent to all the local subscribers, in which case transactions are not distributed across the network.}
\item{To verify the transaction, the Trinity instance that receives the transaction would inject the \emph{unverified} transaction to the blockchain network using Trinity's blockchain APIs (discussed in Section~\ref{sec:blockchain_net}).}
\item{All the Trinity validators will receive the transaction.}
\item{When all the Trinity instances in the network get the \emph{unverified} transaction, they execute the smart contract associated with the topic to check whether the state of the good is adhering to the conditions defined in the smart contract. The outcome of the smart contract execution is shared among the Trinity instances to initiate the consensus process.}
\item{Following a consensus algorithm, the Trinity instances estimate the state of the good. When a majority of Trinity validators approve the transaction, the transaction is entered into a block.}
\item{The transactions approved by the Trinity instances are entered into a new block. The completion of the block creation process denotes the successful replication of the state among all the stakeholders' Trinity instance. The blockchain stores the state of the good in an immutable ledger.}
\item{The \emph{verified} transactions are then sent to the Trinity broker instances, which then relay the \emph{verified} and recorded data to all the subscribers.}
\end{itemize}

The building blocks of Trinity are discussed in the rest of the section.

\subsection{Trinity Broker}
The Trinity broker provides support for smart contracts and immutable ledger in the form of a blockchain. Contemporary data brokers for the IoT forward the published data to all the subscribers without any verification, whereas the Trinity framework verifies and records the transaction on a ledger before forwarding the information to the subscribers.

\begin{figure}[t]
\begin{center}
    \centerline{\includegraphics[height=3in,width=3in, keepaspectratio]{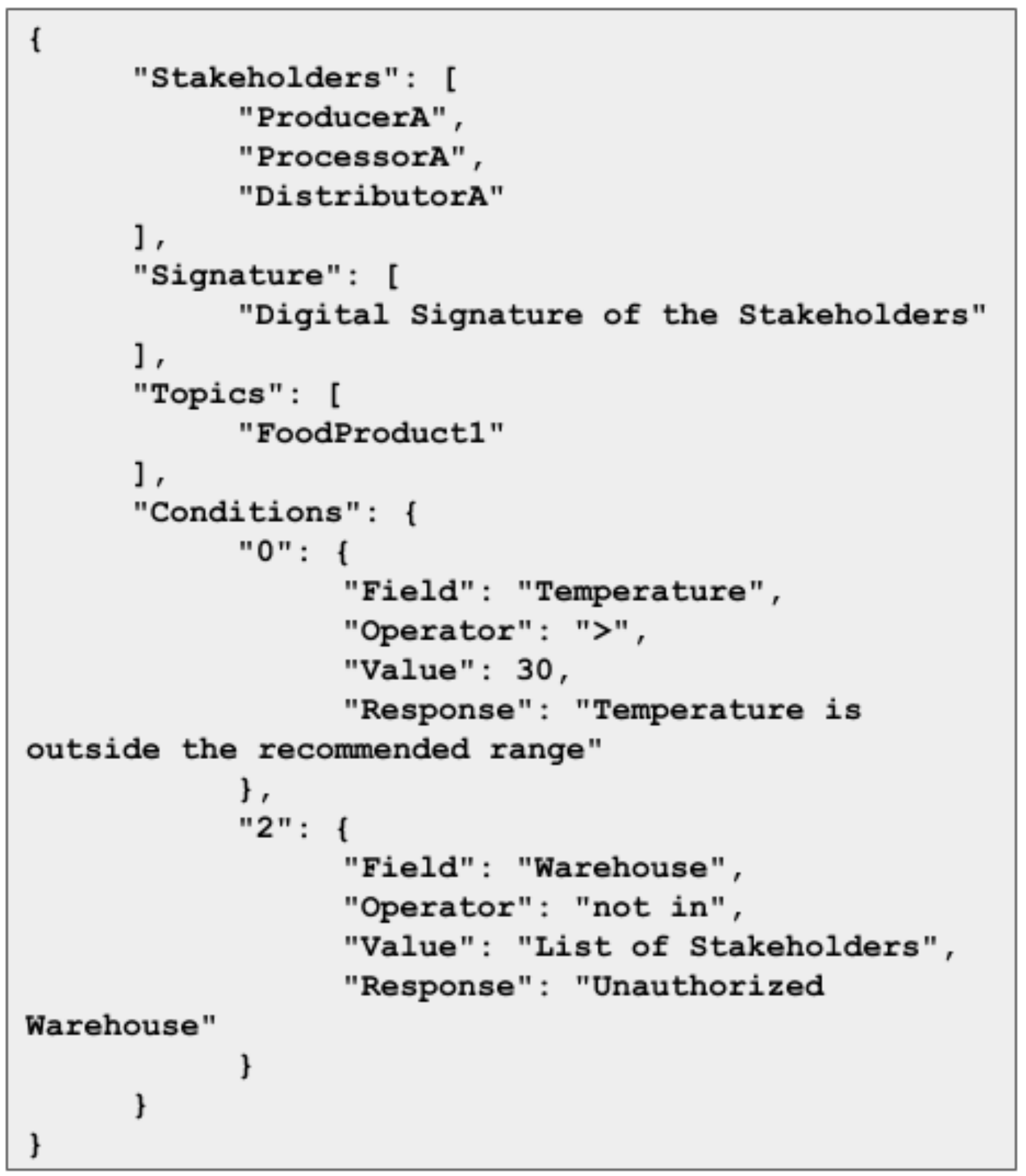} }
    \caption{Example Smart Contract for the Use Case Presented in Figure~\ref{fig:usecase}.}
    \label{fig:contract}
\end{center}
\end{figure}

Our Trinity broker instance consists of a special topic called \emph{Contract}, which allows the Trinity instances to receive a smart contract. Figure~\ref{fig:contract} shows an example smart contract for the use case presented in Figure~\ref{fig:usecase}. Each smart contract must be published under the topic name \emph{Contract}, and it should consist of the list of stakeholders and their digital signatures, topics associated with the contract, and the list of conditions. Upon receiving the contract, the Trinity instance parses and processes the smart contract and replicate the contract among all the Trinity validators in the network. When all the validators receive the copy of the contract, the contract is entered into the blockchain following a consensus protocol. Subsequently, all the data received from the topics registered in the contract are verified and validated by all Trinity validators.

Besides, the Trinity instance allows the stakeholder to monitor the state of the good throughout the supply chain using a reliable infrastructure. This is achieved by returning the \emph{verified} data back to all the brokers. Without the Trinity framework, each stakeholder will have to subscribe to all the brokers in the network and assume that all the organizations are genuine. Moreover, the contemporary brokers send the data from publishers to subscribers without any validation, whereas the Trinity framework executes a smart contract on the data and returns the validated and recorded transaction under a new topic name by appending the \emph{\_verified} to the old topic name. In the Trinity broker instances, all the topics that end with \emph{\_verified} tag are stored in a \emph{immutable ledger}.

\subsection{Trinity Blockchain Network}
\label{sec:blockchain_net}
\begin{figure}[t]
\begin{center}
    \centerline{\includegraphics[height=3in,width=3in,keepaspectratio]{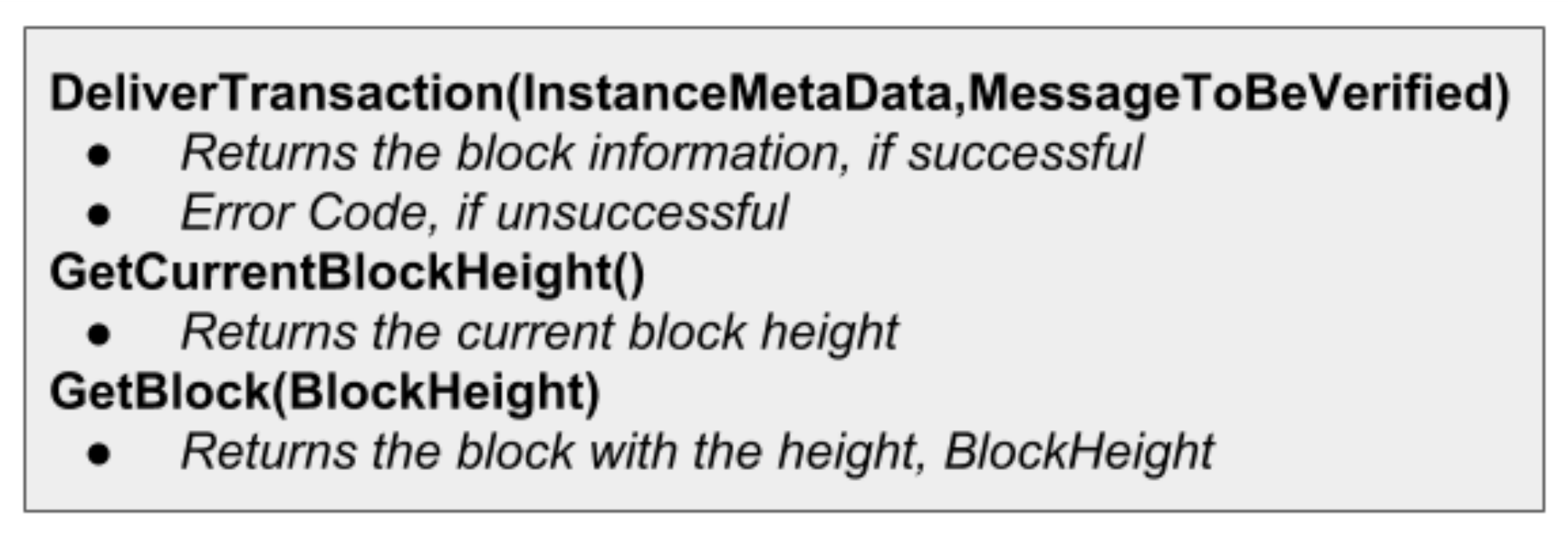} }
    \caption{Trinity APIs to Interact with a Blockchain Network.}
    \label{fig:api}
\end{center}
\end{figure}

The blockchain network is critical for the state replication and data validation. The Trinity framework exposes a set of APIs to interact with a blockchain network. The Trinity broker instance connects with a blockchain network using the APIs listed in Figure~\ref{fig:api}. 

The \emph{DeliverTransaction} API is responsible for state replication, consensus, and the block creation process. The Trinity broker instance sends the information to be verified using the \emph{DeliverTransaction} API along with the necessary instance metadata such as node identifiers and system information to the underlying blockchain framework. The blockchain network replicates the data among the Trinity instances, and everyone in the network executes the smart contract and the consensus algorithm before entering the data into a block. We will discuss the role of the blockchain frameworks below.

The blockchain frameworks typically consist of a consensus protocol, block creation logic, distributed ledger and public-key cryptography. The Trinity framework does not depend on a particular blockchain framework or protocols since the broker interfaces with a blockchain framework via a set of APIs for managing the blockchain-related functionalities. Each Trinity instance has APIs for querying the blockchain network. The \emph{GetCurrentBlockHeight()} API allows the Trinity instance to get the current block height from the blockchain network. Similarly, the \emph{GetBlock(BlockHeight)} API returns the entire block at the height denoted by the \textit{BlockHeight} argument. Although the Trinity broker instances are isolated from the blockchain functionalities, the underlying blockchain framework must have the following components to guarantee immutability:

\textbf{Consensus Algorithm:} The Trinity framework operates on a distributed network owned by multiple organizations. All the authorized members must verify the messages received by the Trinity broker. This verification process relies on a consensus protocol. A system state perceived by one broker in the network must be replicated to other instances of Trinity, and the transaction should be approved by the majority of the devices in the network. Our system can work with consensus protocols such as Proof-of-Work, Proof-of-Stake, or other protocols in the category of Byzantine Fault Tolerance (BFT) such as Tendermint~\cite{buchman2016tendermint} and leader-based protocols such as Raft~\cite{184040}. Note that the resource-consumption and the ability to tolerate device failures depend on the consensus protocol. Our proof-of-concept implementation of the Trinity framework (discussed in Section~\ref{sec:implementation}) used Tendermint, which uses a BFT consensus algorithm, in which at least 2/3 of the devices in the network must authorize a transaction before it can be added to a block. We refer the reader to ~\cite{buchman2016tendermint} for a detailed discussion on the performance and fault tolerance capability of the consensus protocols.

\begin{figure}[t]
\begin{center}
    \centerline{\includegraphics[height=3in,width=3in, keepaspectratio]{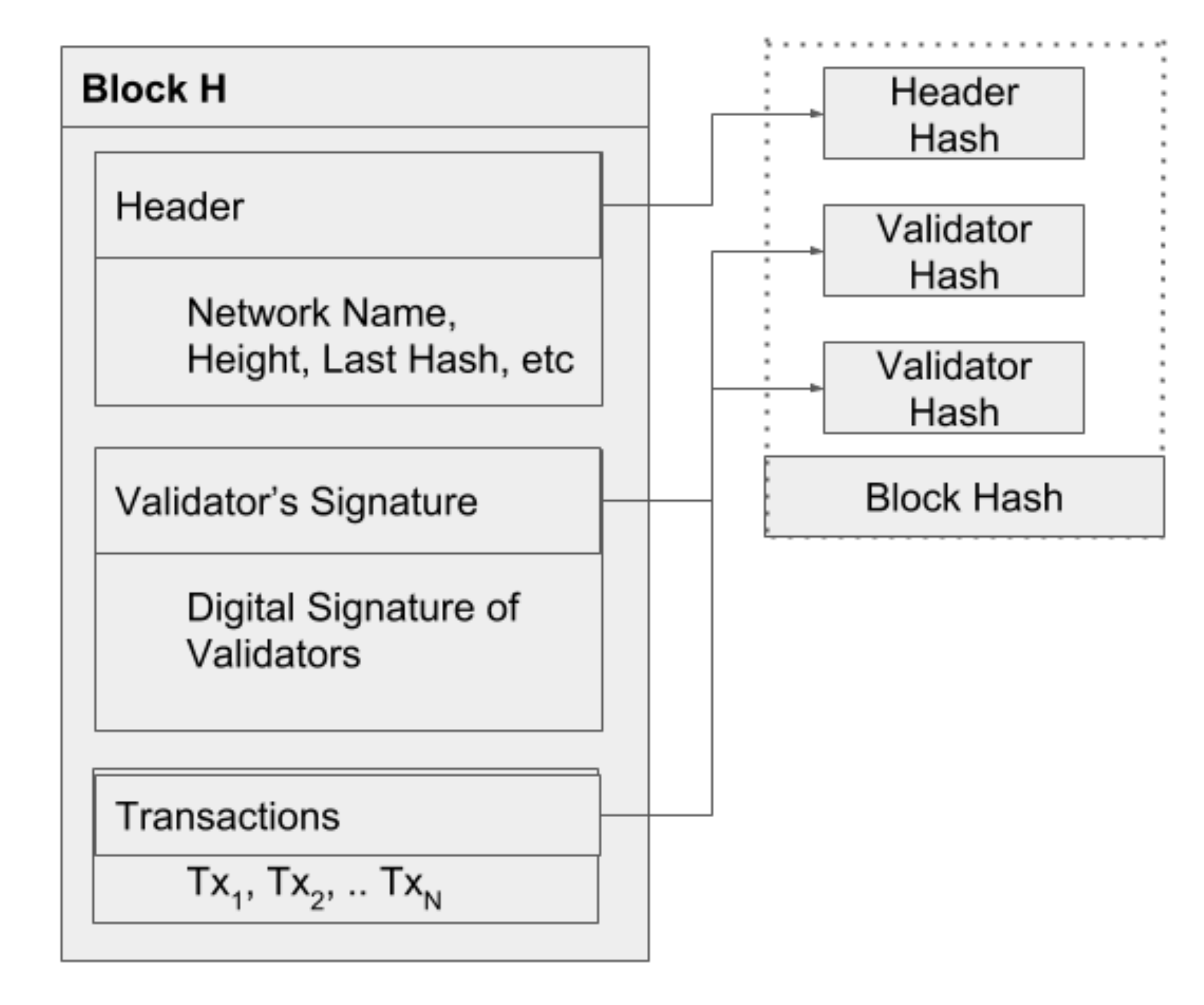} }
    \caption{The structure of a Block~\cite{kwon2014tendermint}.}
    \label{fig:block}
\end{center}
\end{figure}

\textbf{Distributed Ledger:} Contemporary blockchain framework records the transactions in an immutable ledger on all the devices that are involved in the consensus process. Merkel Tree is widely used in the blockchain framework because of its resource-efficiency and low management overhead. Our Trinity requires a ledger mechanism like Merkel Tree for securely storing the transactions. Our proof-of-concept implementation of Trinity (discussed in Section~\ref{sec:implementation}) used Tendermint's Merkle Tree implementation for recording the transaction in the blockchain. Figure~\ref{fig:block} shows the structure of a block. Each block consists of a header, validator's signature, and a list of transactions. The hash of the header, signature, and the list of transactions are hashed again to create the header of a block~\cite{kwon2014tendermint} as shown in Figure~\ref{fig:block}.

\textbf{Public Key Cryptography:} Blockchain framework use the public key cryptography to secure the transactions and to participate in the validation process. Each member of the network create a pair of keys and broadcast their public key to the network to participate in the block creation and consensus process. 

\textbf{Validators:} Validators are devices in the network that are authorized to execute smart contracts, participate in the consensus process, and to create blocks. In the case of the permissioned blockchain, only a designated set of devices can act as the validators, while any capable device can perform validation (and mining) in public blockchains such as BitCoin and Ethereum. We believe that the Trinity framework is better-suited for permissioned blockchains as the parties involved in the transactions are known to the system, and each subscriber will have to register their interest before receiving the data.

In summary, the blockchain-specific tasks of the Trinity framework are loosely coupled with the broker-specific activities, wherein the interaction between the broker and the blockchain network happens via a set of APIs. We believe that this architecture would inspire application developers to replace the publish-subscribe communication model with other IoT protocols to create novel blockchain-based IoT frameworks.

%% file: results.tex
\section{Implementation and Evaluation}
This section discusses the implementation and evaluation of the Trinity framework. Section~\ref{sec:implementation} discusses the implementation details of Trinity. The evaluation setup is introduced in Section~\ref{sec:evalsetup}. The timing overhead of Trinity is presented in Section~\ref{sec:delay}. Section~\ref{sec:network} discusses the network overhead of Trinity. CPU and Memory overhead are presented in Section~\ref{sec:resource}.

\subsection{Implementation of Trinity}
\label{sec:implementation}
We implemented the Trinity framework using Mosquitto (MQTT) Broker~\cite{mqtt} and Tendermint~\cite{buchman2016tendermint} Blockchain framework. MQTT is one of the widely used publish-subscribe communication protocols in the IoT applications. The broker-specific functionalities are implemented on top of MQTT. 

For the blockchain, we used Tendermint~\cite{buchman2016tendermint}, which is an open-source blockchain framework, developed and maintained by a company called \emph{Tendermint}. Tendermint blockchain framework consists of a set of tools for achieving consensus on a distributed network, execution of smart contracts, and creation of blocks. Also, the Tendermint framework isolates the blockchain-related functionalities from the application-specific features, which means any applications can be developed on top of the Tendermint framework, ranging from cryptocurrencies to a distributed chat server. The Tendermint consensus engine allows the application developers to replicate the state of an application across all the Tendermint instances in the network. The state information is fed into the consensus engine using Application Blockchain Interface (ABCI). For the implementation of the Trinity framework, we created a broker application using MQTT~\cite{mqtt} on top of the Tendermint blockchain framework. We used ABCI to bridge the MQTT application with the blockchain framework. The Tendermint framework uses Byzantine Fault Tolerance (BFT) consensus protocol, which means 2/3 of the devices in the network must approve the transactions. When the majority of the devices in the network approve the transaction, the Tendermint framework adds the transaction in a block. All the application-specific software was implemented in NodeJS.

\subsection{Evaluation Setup}
\label{sec:evalsetup}
The Trinity framework was evaluated using a 20-node Raspberry Pi3 network. The Raspberry Pi (Version 3) has ARM Cortex-A53 Quad Core CPU with 1 GB of RAM. We used Hypriot Operating System for the evaluation. All the devices were connected through a LAN. Each data point in the evaluation results is collected by publishing 1000 messages to one of the MQTT brokers in the network. Upon receiving the message, the broker may choose to relay the message to the subscribers as in contemporary publish-subscribe systems or deliver the message to the blockchain framework for validation. The former is denoted as \emph{loopback} in Figure~\ref{fig:e2e_same} and Figure~\ref{fig:e2e_diff}, while the later is referred to as \emph{committed} (validated).

Our evaluation was carried out on 5, 10, 15, and 20 nodes to compare the network performance and end-to-end delay with the scale. We used the default configuration of the Tendermint framework for the evaluation~\cite{kwon2014tendermint}.

\subsection{End-to-End Delay}
\label{sec:delay}
\begin{figure}
    \centering
    \includegraphics[width=0.49\textwidth]{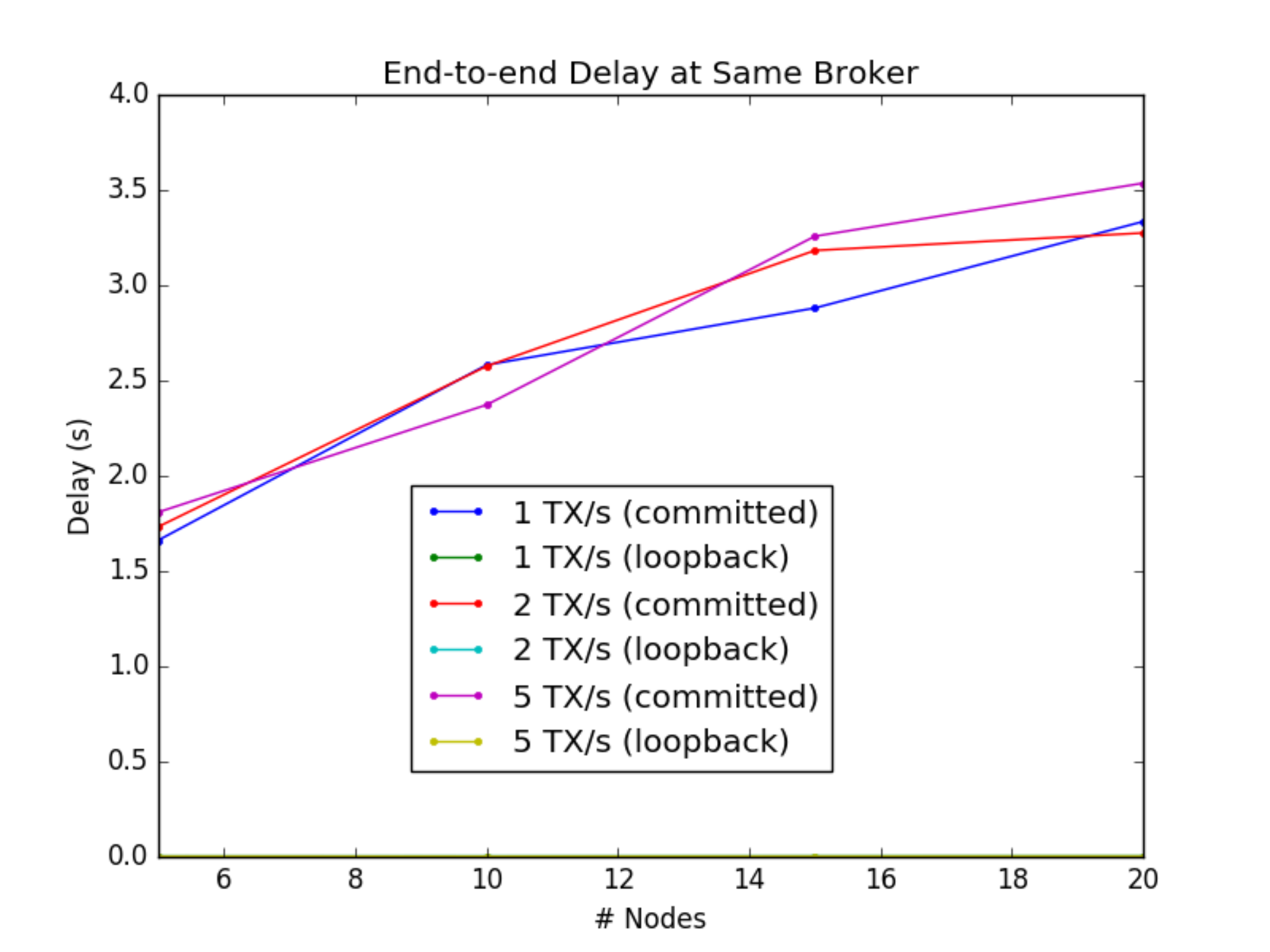}
    \caption{Overhead when subscribed to local broker}
    \label{fig:e2e_same}
\end{figure}

\begin{figure}
    \centering
    \includegraphics[width=0.49\textwidth]{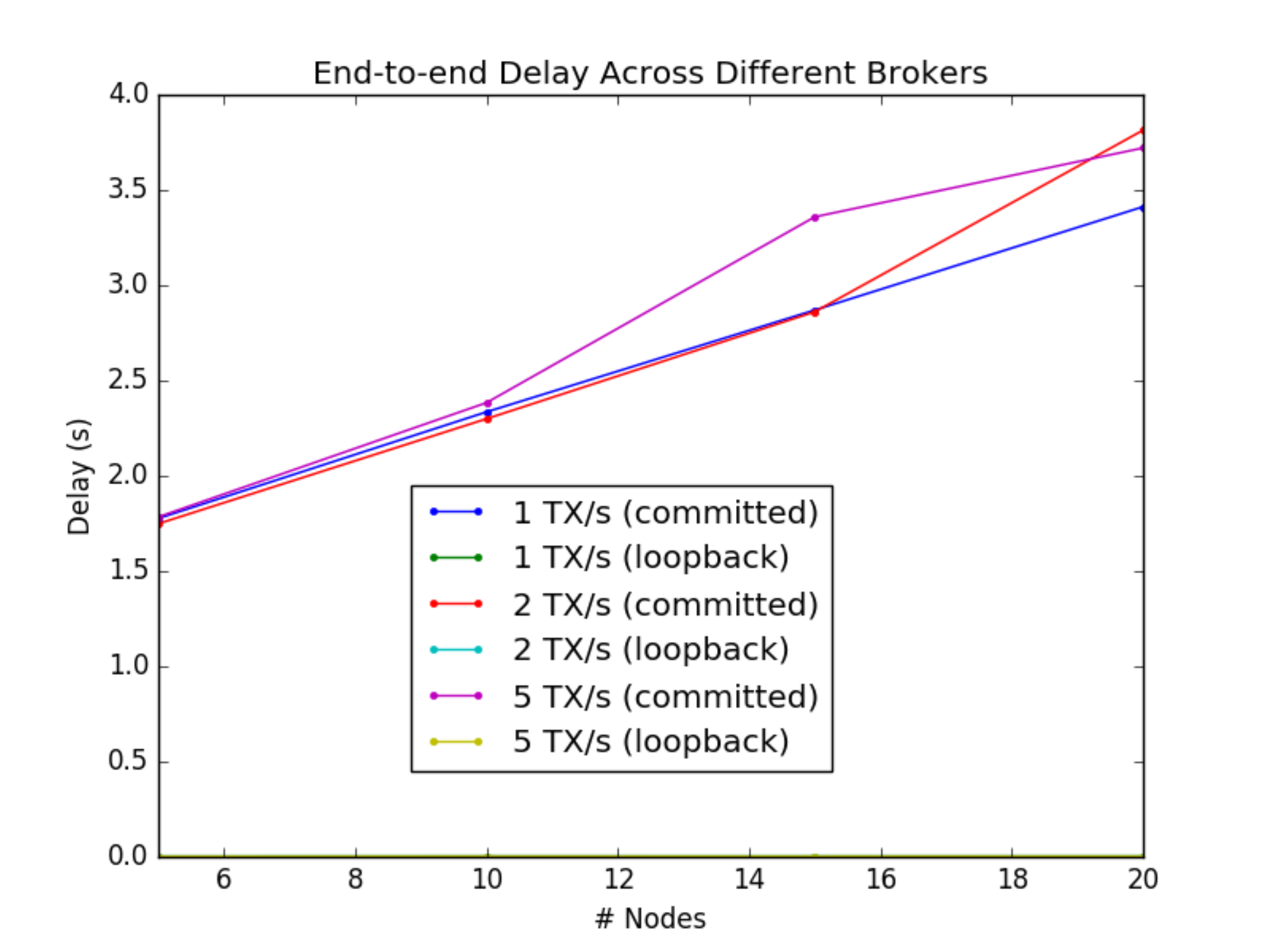}
    \caption{Overhead when subscribed to non-local broker}
    \label{fig:e2e_diff}
\end{figure}

Unlike traditional brokers, the Trinity broker verifies the published message using the smart contract on a blockchain framework and records the transactions in a distributed immutable ledger, which means the subscribers of the Trinity broker gets the recorded and validated transactions. The validation process adds a delay since the Trinity validators must execute the smart contract, consensus algorithm, and the block creation protocols. The subscriber gets the published data after a delay due to the processing overhead of the consensus protocol, smart contract execution, and the block creation. Figure~\ref{fig:e2e_same} and Figure~\ref{fig:e2e_diff} shows the timing overhead of the Trinity framework when publishing to and subscribing from the same broker and publishing to one broker and subscribing for a same verified topic from a different broker respectively.

We measured the end-to-end delay with a publisher sending data every 0.2S, 0.5S, and 1S, which translates to 5 transactions per seconds (TPS), 2 TPS, and 1 TPS respectively. The end-to-end delay increases with the number of devices in the network due to the increase in the number of validators participating in the consensus process. 

The Trinity framework distributes the verified data to all the brokers in the network. All the brokers will have the verified data after the data is added to the immutable ledger. Figure~\ref{fig:e2e_same} shows the timing overhead when publishing to and subscribing from the same broker. The maximum delay is roughly 3.5S for 20 nodes with 5 TPS, and the delay for \emph{loopback} transaction is negligible (approximately 90 milliseconds). When subscribing to the verified from a different broker, the maximum delay increases from 3.5S to 3.7S. From Figure~\ref{fig:e2e_same} and Figure~\ref{fig:e2e_diff}, it is clear that the blockchain-based immutability and verification increases the end-to-end delay, but we believe that this cost outweighs the trust and security benefits.

\subsection{Network Overhead}
\label{sec:network}
\begin{figure}
    \centering
    \includegraphics[width=0.49\textwidth]{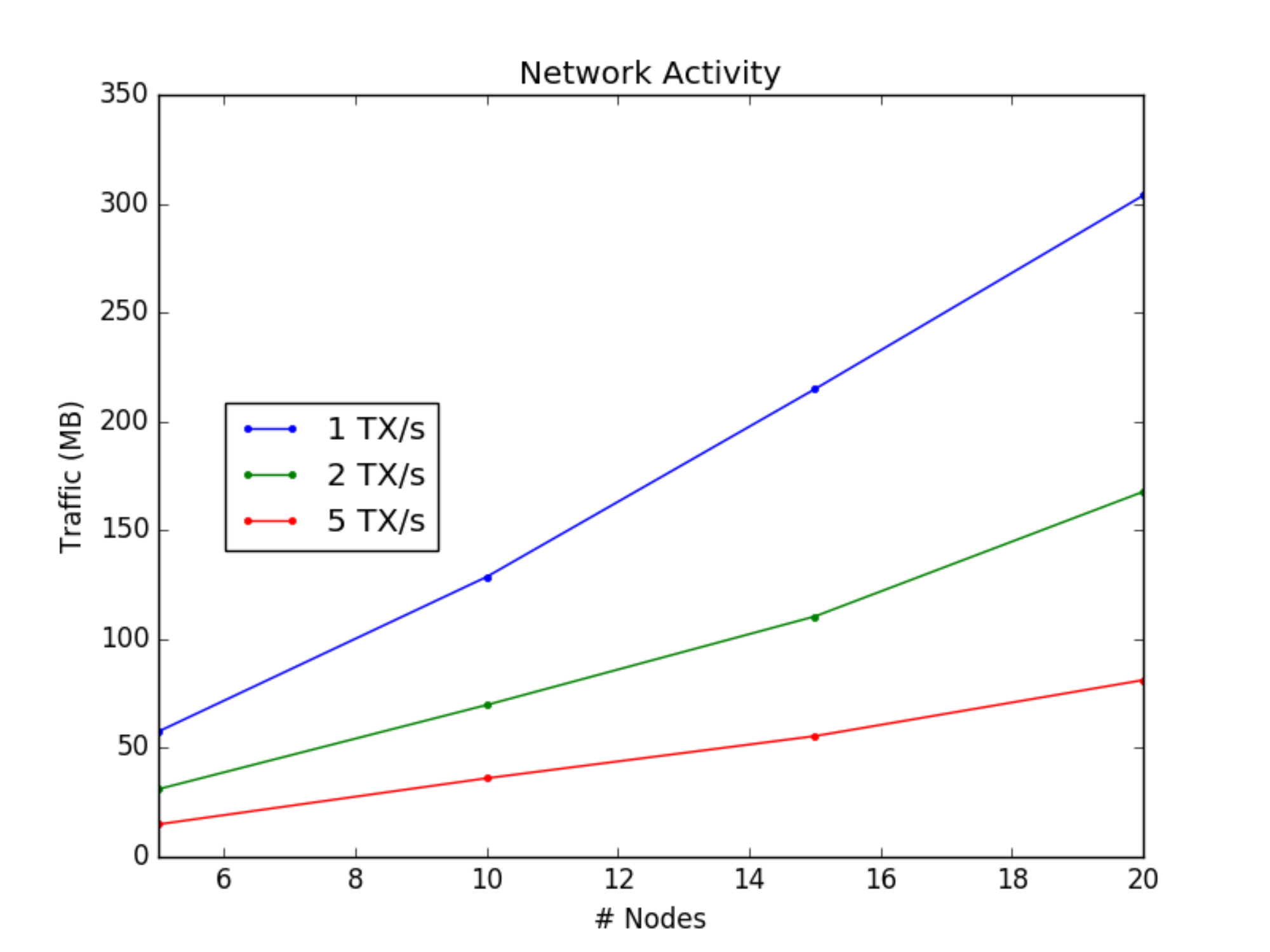}
    \caption{Total network traffic for a single node}
    \label{fig:network_activity}
\end{figure}

The Trinity framework uses a blockchain framework to replicate the state, to execute consensus protocol, and the block creation. All these activities are achieved through the coordination and collaboration of all the devices in the Trinity network. This process generates network traffic. Figure~\ref{fig:network_activity} shows the network overhead of Trinity framework. Note that the implementation of Trinity was carried out on top of the Tendermint framework. The network results presented in Figure~\ref{fig:network_activity} reflects the overhead of the Tendermint framework. 

The networking overhead of the Trinity framework increase as the number of devices in the network increases. Interestingly, the lower the amount of transaction per second, the higher the networking overhead, which is due to the creation of a large number of blocks. The higher TPS typically result in multiple transactions being recorded in a single block, whereas the lower TPS would lead to a single block per transaction. We believe that the maximum network overhead of approximately 300 MB / 1000 Seconds for a 20-device with one TPS is insignificant compared to the benefits offered by the Trinity framework. 

Aggregating multiple transactions can further reduce the networking overhead, but this may come at the cost of high end-to-end delay. We consider such network optimizations as future work.

\subsection{CPU and RAM Usage}
\label{sec:resource}
\begin{figure}
    \centering
    \includegraphics[width=0.49\textwidth]{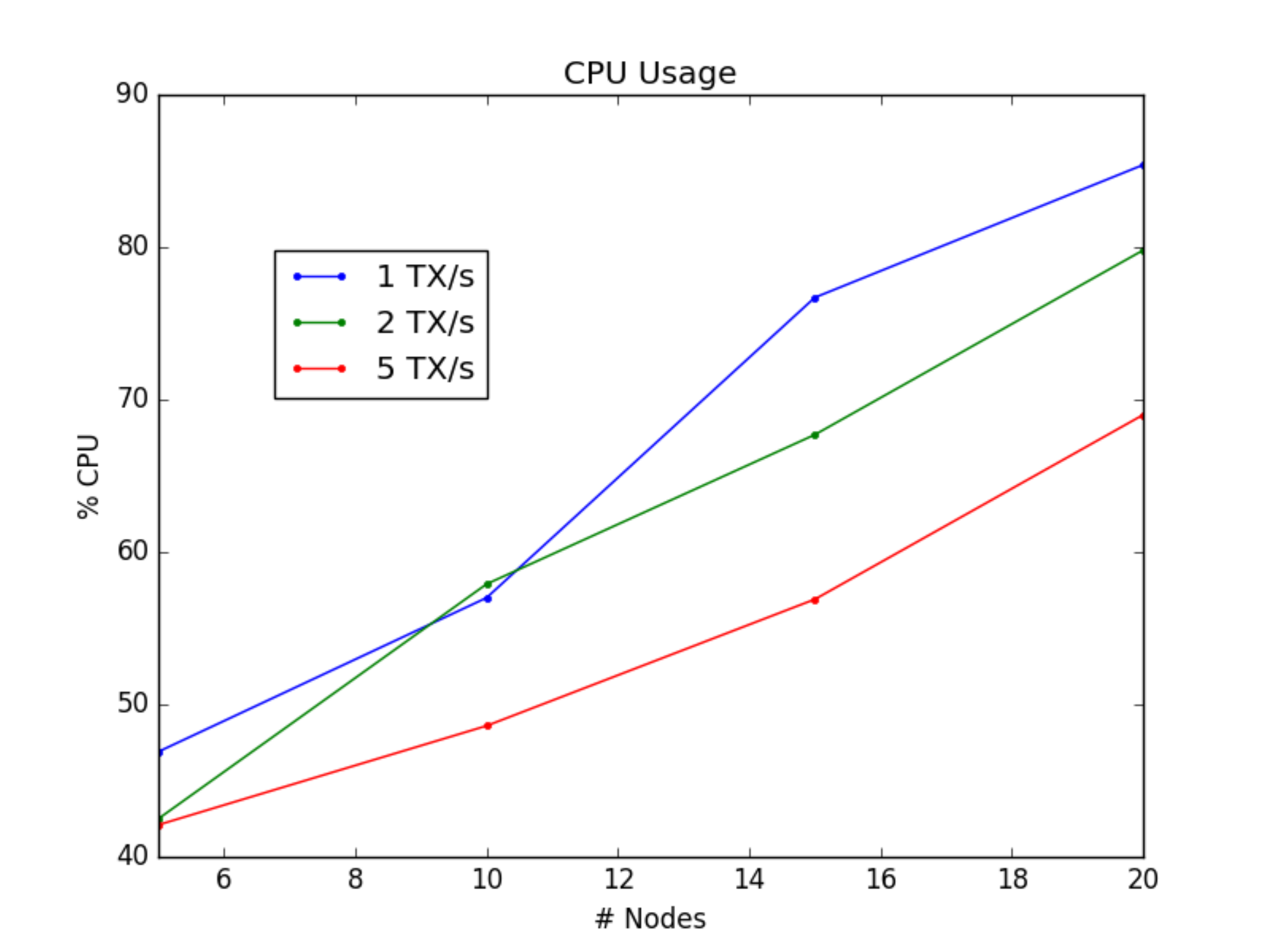}
    \caption{CPU Usage for the consensus and block creation.}
    \label{fig:cpu}
\end{figure}

Figure~\ref{fig:cpu} shows the CPU usage of the Trinity framework. On Raspberry Pi 3 platform, the Trinity framework uses approximately 85\% of the CPU when publishing 1000 transactions at a rate of 1 TPS in 20-node network, since the framework executes the smart contract, consensus algorithm, and block creation protocol every second. Similarly, the maximum RAM usage is measured in a 20-node network when publishing 1000 transactions at a rate of 1 TPS over 15 minutes time interval. The resource overhead of the Trinity framework increases with the number of blocks. By adding multiple transactions in a single block, the CPU and Memory overhead can be minimized. We will investigate the relationship between the number of transactions and the transactions per block using different configurations in our future work.

\begin{figure}
    \centering
    \includegraphics[width=0.49\textwidth]{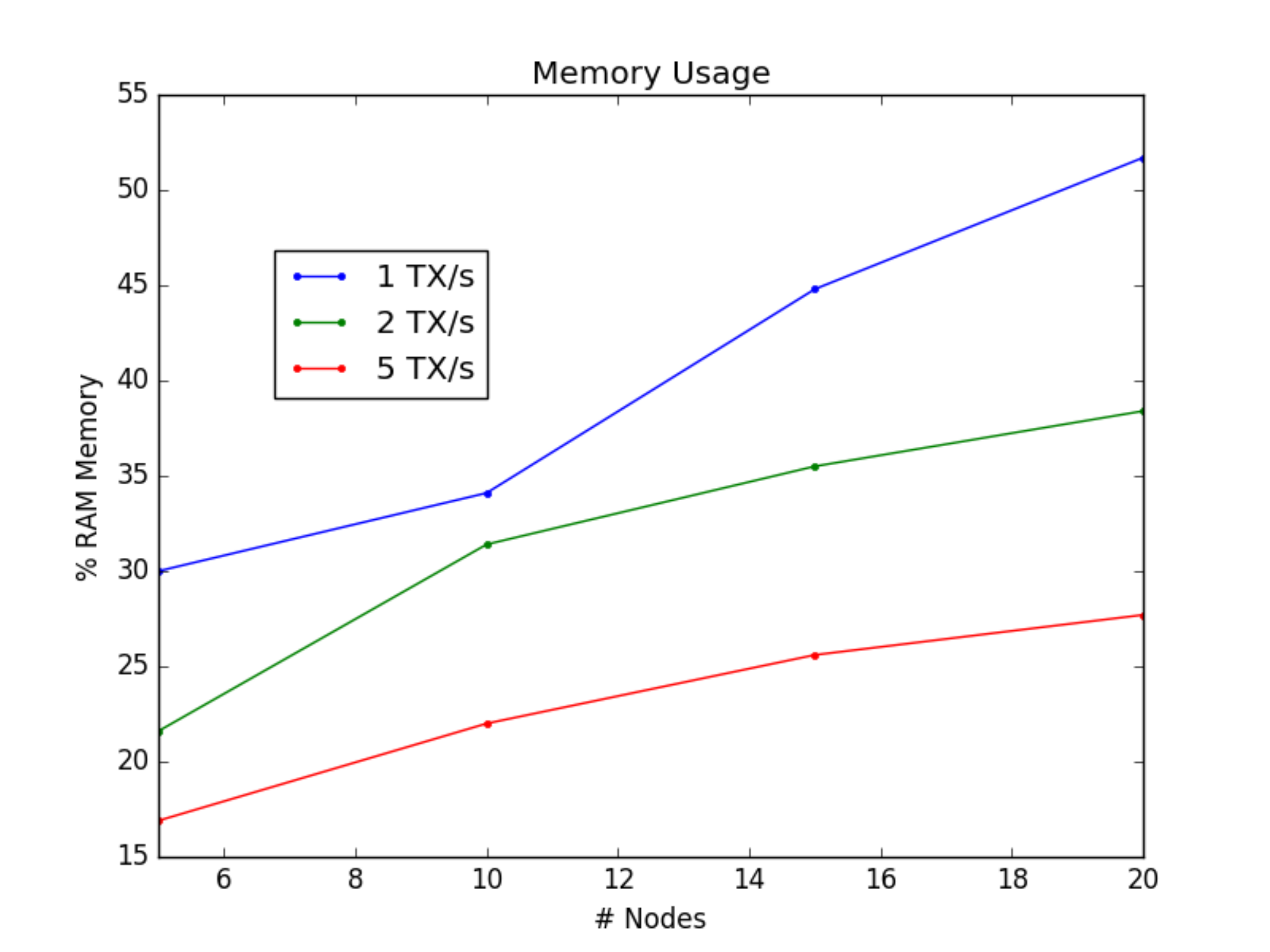}
    \caption{RAM Usage for the consensus and block creation.}
    \label{fig:memory}
\end{figure}

%% file: related_works.tex
Over the past several years, publish-subscribe messaging has proven itself as a dominant messaging paradigm for IoT systems. By decoupling publishers (data sources) from subscribers (data sinks), IoT devices can operate more freely with less prior knowledge of the network they operate in. A variety of publish-subscribe-based software has emerged over the past several years. We discuss them in detail below.

The MQTT protocol \cite{mqtt} is a lightweight publish-subscribe messaging protocol that is popular for IoT systems. Implementations of this protocol can be found in brokers such as Mosquitto \cite{mosquitto}. Mosquitto is primarily designed for use as a single instance, but it has support for bridging, allowing for multiple connected brokers to operate together. Mosquitto's bridging feature effectively duplicates all transmitted messages at every broker, allowing publishers and subscribers to connect to any instance. However, this method of distribution does not guarantee the same ordering of messages at every broker. There is also an implicit assumption of the system being run by a single entity, so there are no trust concerns, and the immutability of data is not provided. We address the issues of ordering and immutability with our blockchain-based broker system.

Kafka \cite{kafka} is a more powerful publish-subscribe broker developed for use in data centers, and it has a rich set of features. Compared to Mosquitto, Kafka is relatively heavyweight and uses a proprietary protocol for communication. It is designed to run in a distributed fashion with built-in support for partitioning and replication. Partitioning is a method for load-balancing across different instances, while replication involves copying the same data across multiple instances. Although Kafka provides ordering guarantees and configurable persistence of messages, it assumes that the software is managed by a single organization, meaning that no trust boundaries are traversed during the operation of the system. There is nothing in place to prevent data tampering as every instance of a Kafka deployment is expected to be owned and operating by a single entity. In our design of Trinity, we do not make this assumption regarding trust. Through the use of a blockchain network, Trinity can guarantee persistence, ordering, and immutability across trust boundaries.

%% file: conclusion.tex
\section{Conclusion}
The blockchain technology has made a significant impact in the world of cryptocurrencies. The building blocks of the blockchain technology such as consensus protocol and distributed ledger are promising for applications beyond cryptocurrencies. In this work, we presented Trinity, which is a publish-subscribe broker with blockchain-based immutability for IoT and supply chain monitoring applications. The Trinity framework increases the trust while performing transactions in applications involving multiple stakeholders such as food supply chain monitoring and smart cities. The Trinity framework decoupled the blockchain-specific tasks from the broker-specific functionalities thereby allowing the application developers to connect their brokers to any blockchain frameworks. The smart contract feature of Trinity automates the data validation process when executing sensitive transactions (IoT data or food products) on infrastructure owned by multiple organizations. The implementation and evaluation of Trinity framework using MQTT and the Tendermint show that the framework can be practically implemented on a contemporary broker and blockchain frameworks. Lastly, the evaluation results on a 20-node Raspberry Pi 3 testbed network shows that the Trinity framework increases the end-to-end delay by approximately 3 seconds while consuming bandwidth and computation resources. We believe that the trust benefits of the Trinity framework outweigh the overhead.

Our future work will extend the smart contract management protocol with either support for a new domain-specific language or adds extensive lists of operators to handle a wide array of conditions and policies. Besides, we will implement the Trinity framework using blockchain frameworks such as HyperLedger Fabric and IOTA to evaluate the performance cost of various consensus algorithms. Lastly, the Trinity framework will be extended with protocols for optimizing computation and networking resources.